\documentclass[aps,prl,nofootinbib,twocolumn,groupedaddress,preprintnumbers,nofootinbib]{revtex4-1}

\usepackage{amssymb}
\usepackage{amsmath}
\usepackage{epsfig}
\usepackage{hyperref}
\usepackage{breakurl}
\usepackage{cleveref}
\usepackage{color}

\makeatletter
\def\simgt{\mathrel{\lower2.5pt\vbox{\lineskip=0pt\baselineskip=0pt
           \hbox{$>$}\hbox{$\sim$}}}}
\def\simlt{\mathrel{\lower2.5pt\vbox{\lineskip=0pt\baselineskip=0pt
           \hbox{$<$}\hbox{$\sim$}}}}
\makeatother

\Crefname{equation}{Eq.}{Eqs.}

\newcommand{\be}{\begin{equation}}
\newcommand{\ee}{\end{equation}}
\newcommand{\bea}{\begin{eqnarray}}
\newcommand{\eea}{\end{eqnarray}}
\newcommand{\eq}[2]{\be\begin{aligned}#1 \label{#2}\end{aligned}\ee}

\newcommand{\Eq}[1]{Eq.~\eqref{#1}}

\def\Section#1{\smallskip \noindent {\bf #1}}

\begin{document}

\title{Geometry-Kinematics Duality}

\author{Clifford Cheung, Andreas Helset, and Julio Parra-Martinez}
\affiliation{Walter Burke Institute for Theoretical Physics,
California Institute of Technology, Pasadena, CA 91125}

\begin{abstract}

We propose a mapping between geometry and kinematics that implies the classical equivalence of any theory of massless bosons---including spin and  exhibiting arbitrary derivative or potential interactions---to a nonlinear sigma model (NLSM) with a momentum-dependent metric in field space.
From this kinematic metric we construct a corresponding kinematic connection, covariant derivative, and curvature, all of which transform appropriately under general field redefinitions, even including derivatives.  We show explicitly how all tree-level on-shell scattering amplitudes of massless bosons are equal to those of the NLSM via the replacement of geometry with kinematics.  Lastly, we describe how the recently introduced geometric soft theorem of the NLSM, which universally encodes all leading and subleading soft scalar theorems, 
also captures the soft photon theorems.

\end{abstract}

\preprint{CALT-TH 2022-006}

\maketitle

\Section{Introduction.}  Color-kinematics duality is an extraordinary structure relating the on-shell scattering amplitudes of gauge theory and gravity \cite{Bern:2008qj,Bern:2010ue,Bern:2019prr}.  Mechanically, the former is mapped to the latter by promoting color structure constants $f^{abc}$ to momentum-dependent functions that conform to identical antisymmetry properties and Jacobi identities.  Hence, gravity is simply a gauge theory for which color has been substituted for kinematics.

In this paper, we argue that the classical dynamics of {\it any theory of massless bosons} is similarly obtained via a mapping of {\it geometry to kinematics}.  Our starting point is the nonlinear sigma model (NLSM), which defines scalar fields as coordinates on an internal field space manifold endowed with a metric and a well-known apparatus of geometrical structures \cite{Meetz:1969as,Honerkamp:1971sh,Honerkamp:1971xtx,Ecker:1971xko,osti_4340109,Alvarez-Gaume:1981exa,Alvarez-Gaume:1981exv,Boulware:1981ns,Howe:1986vm,Dixon:1989fj,Alonso:2015fsp,Alonso:2016oah,Nagai:2019tgi,Helset:2020yio,Cohen:2021ucp}.  The dynamics of the NLSM are then specified by the field space metric and its derivatives at the vacuum, $g_{ij}$, $g_{ij,k}$, $g_{ij,kl}$, etc.  By uplifting these metric coefficients to momentum-dependent functions, we obtain a kinematic metric that defines a generalized field space geometry,  simultaneously encoding flavor, spin, and momentum all at once.  We then construct the kinematic analogs of the connection, covariant derivative, and curvature, which transform covariantly under arbitrary field redefinitions.  From this perspective, all massless bosons---{\it e.g.}~in higher-derivative scalar theories, $\phi^3$ theory, and Yang-Mills (YM) theory---are described by an NLSM for a concrete, calculable choice of momentum-dependent metric. Our results imply the invariance of tree-level on-shell scattering amplitudes under arbitrary field redefinitions, even including derivatives of fields.

A powerful corollary of this geometry-kinematics duality is that any tree-level on-shell scattering amplitude of massless bosons is directly obtained from an NLSM amplitude after a suitable replacement of geometry with kinematics. This rewriting of the amplitude is field-redefinition invariant term by term, though on occasion at the expense of manifest locality.  Furthermore, under this substitution, the geometric soft theorem for any massless scalar \cite{Cheung:2021yog}---which generalizes the Adler zero \cite{Adler:1964um} and dilaton soft theorems \cite{Callan:1970yg, Boels:2015pta, Huang:2015sla, DiVecchia:2015jaq}---also encodes the leading and subleading soft photon theorems \cite{Low:1958sn, Weinberg:1964ew, Weinberg:1965nx,  Burnett:1967km}.

\Section{Geometry of Field Space.}   
The NLSM is a general two-derivative theory of scalar fields $\phi^i$ described by 
\eq{
L &=  -\tfrac{1}{2} g_{ij}(\phi) \partial \phi^i  \partial \phi^j , 
}{L_NLSM}
where the spacetime metric is in mostly plus signature.  The field space metric expanded about the vacuum is
\eq{
g_{ij}(\phi) =  g_{ij} + g_{ij,k} \phi^k +\tfrac{1}{2} g_{ij,kl} \phi^k \phi^l+ \cdots,
}{metric_expand}
where hereafter any quantity without an explicit $\phi$ argument is to be evaluated at the vacuum.

On-shell scattering amplitudes are invariant under changes of field basis whereby $\phi^i$ is rewritten in terms of $\tilde \phi^i$ through the mapping $\phi^i(\tilde \phi)$.
The Lagrangian becomes $L =  -\tfrac{1}{2} \tilde g_{ij}(\tilde \phi) \partial \tilde \phi^i  \partial \tilde \phi^j$, so the metric is a tensor that transforms as
$\tilde g_{ij}(\tilde\phi) = \tfrac{\partial \phi^k }{\partial \tilde \phi^i}   \tfrac{\partial \phi^l}{\partial \tilde \phi^j} g_{kl}(\phi)$.  In turn, $g_{ij}(\phi)$ is associated with a menagerie of geometric objects, such as the connection $\Gamma_{ijk}(\phi)$, curvature $R_{ijkl}(\phi)$, and so on.    

Since on-shell scattering amplitudes of the NLSM are field basis independent, they depend solely on geometric invariants, so {\it e.g.}~at four- and five-point they are \cite{Cheung:2021yog}
\eq{
A_{ijkl} =&\phantom{{}+{}} R_{ijkl} u + R_{ikjl}  s ,\\
A_{ijklm} =&\phantom{{}+{}} \nabla_k R_{iljm} s_{45} +  \nabla_l R_{ikjm} s_{35} + \nabla_l R_{ijkm} s_{25} \\
&+ \nabla_m R_{ikjl} s_{34} + \nabla_{m} R_{ijkl} (s_{24} + s_{45}) ,
}{amplitudes}
where $s_{ij} = -(p_i +p_j)^2$ and $s,t,u=s_{12},s_{23},s_{31}$.
Hereafter, indices in lexographical order $i,j,k,$ etc., will denote the flavors of the corresponding legs, $1,2,3,$ etc.

\Section{Geometry of Kinematics.} Consider a single scalar with arbitrary interactions.   Despite the absence of flavor, there exists a notion of kinematic geometry in which {\it momentum} plays the role of the index.  Concretely, we map the flavor multiplet of the NLSM to a single scalar,
\eq{
\phi^i  \quad \rightarrow \quad \phi(p).
}{map_field}
Under this replacement, the internal field space metric is mapped to the momentum-dependent kinematic metric,
\eq{
g_{ij} \quad &\rightarrow \quad g(p_1, p_2) \delta(p_{12})\\
g_{ij,k} \quad &\rightarrow \quad g(p_1, p_2 |p_3) \delta(p_{123})\\
g_{ij,kl} \quad &\rightarrow \quad g(p_1, p_2| p_3,p_4)  \delta(p_{1234}), 
}{map_metric}
and so on, where $p_{1\cdots n} = p_1 + \cdots + p_n$. Note that the only constraint on the functions on the right-hand sides is that they are separately permutation invariant under exchange of 1 and 2, and under exchange of $3,4,\cdots, n$.  This ensures that the kinematic metric has the same symmetry properties as the usual metric on field space.

For later convenience, we canonically normalize the kinetic terms.  This condition in the NLSM maps to 
$g_{ij} = \delta_{ij}  \rightarrow  g(p_1, p_2) = 1$.
Meanwhile, any sum over an internal field space index maps to an integral over momentum,
$\sum_i    \rightarrow  \int_{p} = \int \tfrac{ d^Dp}{(2\pi)^D}$.
Together, these imply that the field with lowered indices maps to $\phi_i = g_{ij} \phi^j  \rightarrow \phi(-p)$,
so raised and lowered indices correspond to incoming and outgoing momenta. We will refer to  the above replacements as the geometry-kinematics mapping or duality.  

Applying this replacement to the {\it Lagrangian} for the NLSM in \Eq{L_NLSM}, we obtain the {\it action} for a single scalar expressed in momentum space, 
\begin{widetext}
\eq{
S =  \tfrac{1}{2} \int\displaylimits_{p_1, p_2}  (p_1 \cdot p_2) \phi(p_1) \phi(p_2) \bigg[  &  \delta(p_{12}) + \int\displaylimits_{p_3}  g(p_1, p_2 |p_3) \delta(p_{123})  \phi(p_3)  + \tfrac{1}{2}\int\displaylimits_{p_3,p_4}g(p_1, p_2| p_3, p_4)  \phi(p_3) \phi(p_4)  \delta(p_{1234}) +\cdots 
\bigg],
}{action}
\end{widetext}
which actually has sufficient freedom to parameterize arbitrary higher-derivative and potential interactions.  As we will see, one simply compares the action for a given scalar theory directly to \Eq{action}, reading off the kinematic metric coefficients in \Eq{map_metric} by inspection.  We will implement this procedure in numerous examples later on.

By design, this setup reduces to the usual NLSM whenever the metric coefficients in \Eq{map_metric} are constant.  
Furthermore, from \Eq{map_metric} we see that geometry-kinematics duality maps the full metric $g_{ij}(\phi)$ in \Eq{metric_expand}, which is a {\it function} of the fields, to a kinematic metric that is a {\it functional} of the fields.

\Section{Invariants and Field Redefinitions.}   To calculate the kinematic analog of a given geometric invariant, we express the invariant in the NLSM in terms of the field space metric and its derivatives and then apply the geometry-kinematics replacement. For example, the NLSM connection is
$\Gamma_{ijk} = \tfrac{1}{2} \big[ g_{ij,k}+ g_{ik,j}-g_{jk,i} \big]$, which maps to $\Gamma_{ijk}  \rightarrow  \Gamma(p_1, p_2 ,p_3) \delta(p_{123})$ for
\eq{
\Gamma(p_1, p_2 ,p_3) =\tfrac{1}{2} \big[ g(p_1, p_2 |p_3) +g(p_1, p_3 |p_2)-g(p_2, p_3 |p_1)  \big] .
}{}
Similarly, the NLSM curvature is expressed in terms of the metric as $R_{ijkl} = \tfrac{1}{2}[g_{il,jk}+g_{jk,il} - g_{ik,jl}  -g_{jl,ik}] +\cdots$, which maps to $R_{ijkl}  \rightarrow  R(p_1, p_2, p_3,p_4)  \delta(p_{1234}) $, where
\eq{
R(p_1,p_2,p_3,p_4) = \tfrac{1}{2} \big[ g(p_1, p_4 |p_2,p_3) + g(p_2, p_3 |p_1,p_4) \cdots \big].
}{Riemann_kinematic}
Here the sums over internal indices map to momentum integrals, so for example a term like $g_{ik,m}g^{mn} g_{nj,l} $ is sent to $
g(p_1, p_3|-p_{13}) g( -p_{24},p_2|p_4) \delta(p_{1234})$,
and so on.  Higher-point geometric objects like $\nabla_i R_{jklm}$ are similarly calculated by expanding explicitly in terms of the metric and then applying the geometry-kinematics replacement.

The kinematic curvature in \Eq{Riemann_kinematic} transforms as a tensor---even for {\it off-shell} momenta---under the kinematic generalization of a change of coordinates.  Indeed, the geometry-kinematics replacement sends an arbitrary NLSM field redefinition, $\phi^i = \tilde \phi^i  +\tfrac{1}{2} c^i_{\phantom{i} jk}\tilde \phi^j \tilde \phi^k +\cdots $, to
\eq{
\phi(p_1) = \tilde \phi(p_1) &+ \tfrac{1}{2} \int\displaylimits_{p_2} c(p_1, -p_2,p_{12}) \tilde \phi(p_2) \tilde \phi(-p_{12}) +\cdots ,
}{transform_field_kinematic}
where the functions on the right-hand side are arbitrary.  The NLSM  Jacobian, $\tfrac{\partial \phi^i}{ \partial \tilde \phi^j} = \delta^i_j  + c^i_{\phantom{i} jk} \tilde \phi^k +\cdots$, maps to
\eq{
\tfrac{\partial \phi(p_1)}{\partial \tilde \phi(p_2)} = \delta(p_1-p_2)+c(p_1, -p_2,p_{12}) \tilde \phi(-p_{12})  +\cdots .
}{}
The astute reader will notice that the transformation in \Eq{transform_field_kinematic} is literally the momentum-space expression of a field redefinition of the single scalar allowing for {\it derivatives}.  For example, the specific transformation
$\phi = \tilde \phi +  \tfrac{1}{2} c_3 \partial \tilde \phi \partial \tilde \phi +  \tfrac{1}{6}c_4 \tilde \phi \partial \tilde \phi \partial \tilde \phi  + \cdots$
is obtained from
$c(p_1,p_2,p_3) = -c_3 (p_2 \cdot p_3)$,
$c(p_1,p_2,p_3,p_4) = -\tfrac{1}{3} c_4 [(p_2 \cdot p_3)+(p_3 \cdot p_4)+(p_4 \cdot p_2)] $, and so on. 
The kinematic curvature is a tensor under an arbitrary derivative field redefinition, {\it i.e.}~for any choice of functions in \Eq{transform_field_kinematic}.

In summary, kinematic geometry is formally identical to the standard geometry in the NLSM, albeit with regular derivatives with respect to a constant scalar background effectively replaced with functional derivatives.  

\Section{Scalar Theory.} Let us now examine some examples, starting with the case of a single scalar field.

\smallskip
\noindent {\it i) Nambu-Goldstone Boson Theory.} Consider a derivatively coupled scalar field described by the Lagrangian, 
\eq{
L 
&=  - \tfrac{1}{2} \partial \phi \partial \phi \big[ 1-  \tfrac{\lambda_4}{4}\partial \phi \partial \phi +  \tfrac{\lambda_6}{8}(\partial \phi \partial \phi)^2 +  \cdots \big],
}{L_dphi4}
where $\lambda_{4} = \lambda_{6}=1$ corresponds to Dirac-Born-Infeld theory.  \Eq{L_dphi4} matches \Eq{action} for the choice of metric,
\eq{
g(p_1, p_2|p_3,p_4) &=\tfrac{\lambda_4}{2} (p_3 \cdot p_4) \\
g(p_1, p_2|p_3,p_4,p_5,p_6) &={\scriptstyle \lambda_6}[ (p_3 \cdot p_4 )(p_5\cdot p_6) + \cdots].
}{}
The corresponding kinematic curvature is
\eq{
R(p_1,p_2,p_3,p_4) &=- \tfrac{\lambda_4}{4} (t-u),
}{Riemann_dphi4}
evaluated here at four-point on-shell kinematics.   As required, this object has the requisite symmetry properties, $R(p_1,p_2,p_3,p_4) = -R(p_2,p_1,p_3,p_4) = R(p_3,p_4,p_1,p_2)$ and satisfies the first Bianchi identity, $R(p_1,p_2,p_3,p_4)+ R(p_2,p_3,p_1,p_4) +R(p_3,p_1,p_2,p_4)=0$.  

Plugging \Eq{Riemann_dphi4} into the four-point amplitude for the NLSM in \Eq{amplitudes}, we obtain the four-point amplitude
\eq{
A(p_1,p_2,p_3,p_4) = \tfrac{\lambda_4}{4} (s^2 + t^2 + u^2),
}{}
which exactly matches the answer computed directly from \Eq{L_dphi4}.  By computing the six-point kinematic geometric invariants---which automatically satisfy the first and second Bianchi identities---and inserting them into the corresponding NLSM six-point amplitude in \cite{Cheung:2021yog}, we also obtain the correct six-point answer.

We have also verified that the four- and six-point kinematic curvature invariants are invariant under derivative field redefinitions, even for off-shell kinematics.  So while NLSM amplitudes must be on-shell in order to be expressed solely in terms of curvature invariants, as in \Eq{amplitudes}, the curvature invariants themselves can be evaluated off-shell, and as advertised,  they still satisfy all properties required of true geometric invariants.

\smallskip
\noindent {\it ii) Higher-Derivative Theory.} Our analysis generalizes to theories with even higher-derivative interactions.  For example, consider such a theory defined by
\begin{align}
\label{L_SG}
L= &  - \tfrac{1}{2} \partial \phi \partial \phi \big[ 1- \tfrac{\lambda_4}{6}([\Pi]^2- [\Pi^2])  \\ \nonumber
& +\tfrac{\lambda_6}{120} ([\Pi]^4 - 6[\Pi]^2  [\Pi^2]+3 [\Pi^2]^2+8[\Pi][\Pi^3]-6 [\Pi^4]) \big],
\end{align}
where $\Pi_{\mu\nu} = \partial_\mu \partial_\nu \phi$ and the brackets indicate traces over Lorentz indices.  Here $\lambda_{4} = \lambda_{6}=1$ defines the special Galileon theory \cite{Cheung:2014dqa, Hinterbichler:2015pqa}.  The kinematic metric is
\eq{
g(p_1, p_2|p_3,p_4) &=\tfrac{\lambda_4}{3} (p_3 \cdot p_4)^2,
}{}
and kinematic curvature for on-shell kinematics is
\eq{
R(p_1,p_2,p_3,p_4) &= \tfrac{\lambda_4}{12} (t^2-u^2),
}{Riemann_SG}
which again satisfies all appropriate symmetry properties.  Inserting the kinematic curvature into \Eq{amplitudes} yields 
\eq{
A (p_1,p_2,p_3,p_4) &= -\tfrac{\lambda_4}{4} stu,
}{}
which is the correct four-point amplitude.

In order to define the kinematic geometry of a theory, a choice must be made for the initial metric. Curiously, this  choice is in general ambiguous.  For example, consider the six-point operator $ (\partial \phi)^2 [\Pi^2]^2$ in \Eq{L_SG}, which is obtained from \Eq{action} for the kinematic metric
\eq{
g(p_1,p_2|p_3,p_4,p_5,p_6) \sim (p_3\cdot p_4)^2 (p_5\cdot p_6)^2 + \cdots,
}{}
or, alternatively, for the choice
\eq{
g'(p_1,p_2|p_3,p_4,p_5,p_6) \sim  (p_1\cdot p_2) \left[ (p_3\cdot p_4)^2 (p_5\cdot p_6) + \cdots \right] .
}{}
Inserted into \Eq{action}, $g$ and $g'$ produce the same Lagrangian.  Their difference arises only because in each metric we have chosen to associate a different pair of fields to the momenta $p_1$ and $p_2$ in \Eq{action}.  Strangely, $g$ and $g'$ induce curvature six-point invariants, $\nabla^2 R$ and $\nabla'^2 R'$, which are {\it not equal}.  Despite this, both choices reproduce the correct on-shell six-point amplitude.  Furthermore, applying a field redefinition to either $g$ or $g'$ leaves their respective curvature invariants $\nabla^2 R$ and $\nabla'^2 R'$ unchanged, albeit still unequal to each other.   Yet another ambiguity arises from total derivatives, which evaporate from observables but can still contribute to the kinematic metric.  Let us return to these issues shortly.

\smallskip \noindent {\it iii) Potential Theory.}   
Rather surprisingly, our formalism can be applied to a theory with an arbitrary potential even though the NLSM has intrinsic derivative interactions.  In particular, observe that the theory
\eq{
L =  -\tfrac{1}{2} \partial \phi \partial \phi - V(\phi),
}{L_V}
corresponds to \Eq{action} with the kinematic metric,
\eq{
 g(p_1, p_2|p_3,\cdots, p_n) &= -\tfrac{2}{p_1\cdot p_2} \tfrac{(n-2)!}{n!}{\scriptstyle V_{(n)}},
}{metric_V}
where $V_{(n)}$ is the $n$th derivative of the potential.  The factors of $\tfrac{1}{p_1\cdot p_2}$ have the sole purpose of canceling the momenta that appear in the very definition of the NLSM.  The corresponding kinematic curvature is
\eq{
R(p_1,p_2,p_3,p_4) &= {\scriptstyle V_{(3)}^2} \left( \tfrac{1}{t^2}- \tfrac{1}{u^2} \right) +\tfrac{V_{(4)}}{3}  \left( \tfrac{1}{t}- \tfrac{1}{u} \right),
}{Riemann_potential}
which inserted into \Eq{amplitudes} correctly yields
\eq{
A(p_1,p_2,p_3,p_4) &= -{\scriptstyle V_{(3)}^2} \left( \tfrac{1}{s}+\tfrac{1}{t} + \tfrac{1}{u} \right) - {\scriptstyle V_{(4)}} .
}{}
We have verified explicitly that the resulting five- and six-point amplitudes are also correct.

The above result is somewhat miraculous because the amplitudes in the NLSM have a {\it strict subset} of the factorization channels that appear in the theory defined in \Eq{L_V}.  
In particular, the NLSM amplitudes have no factorization channels involving three-point subdiagrams. Nevertheless, the non-local momentum dependence in \Eq{metric_V} enters precisely in a way that automatically regenerates all three-point factorization channels.

Note that the on-shell three-point amplitude itself is slightly subtle in this construction---it technically vanishes in the NLSM since $A_{ijk} \sim g_{ij,k} (p_1 \cdot p_2) +\cdots$ and all dot products of momenta vanish for three-point on-shell kinematics.  However,  the metric in \Eq{metric_V} has a denominator that similarly vanishes in such a way that the resulting three-point amplitude is actually non-zero.

\smallskip \noindent {\it iv) Arbitrary Theory.}   There exist some features universal to any massless scalar theory.  For example, observe that \Eq{Riemann_dphi4}, \Eq{Riemann_SG}, and \Eq{Riemann_potential} take the form
\eq{
R(p_1,p_2,p_3,p_4) \sim t^\rho - u^\rho,
}{}
where  $\rho$ is the parameter defined in \cite{Cheung:2014dqa,Cheung:2015ota,Cheung:2016drk} which quantifies the numbers of derivatives per interaction. The above expression for the kinematic curvature is actually universal, since it is the unique function with the appropriate symmetries that satisfies the Bianchi identity.

For any massless scalar theory it is also possible to explicitly construct a kinematic metric directly from its Feynman rules.  Inspired by \Eq{metric_V}, we define
\eq{
 g(p_1, p_2|p_3,\cdots, p_n) &= \tfrac{2}{p_1\cdot p_2} \tfrac{(n-2)!}{n!}{F(p_1, p_2, p_3, \cdots, p_n)},
}{metric_from_Feynman}
where $F$ is the $n$-point Feynman vertex.  This choice of kinematic metric is properly permutation invariant and eliminates the ambiguities arising from metric choice and total derivatives mentioned previously.  

Amusingly, there is an alternative choice of coordinates---the kinematic analog of Riemann normal coordinates---defined through the on-shell scattering amplitudes themselves.  In particular, consider
\eq{
g(p_1,p_2|p_3)&=0 \\ 
 g(p_1, p_2|p_3, p_4) &= \tfrac{1}{6(p_1\cdot p_2)} A(p_1, p_2, p_3, p_4)\\
 g(p_1, p_2|p_3, p_4,p_5) &= \tfrac{1}{10(p_1\cdot p_2)} A(p_1, p_2, p_3, p_4,p_5),
}{}
which generates kinematic curvature invariants such as
\eq{
R(p_1, p_2,p_3, p_4) &= \tfrac{1}{3} (\tfrac{1}{u}-\tfrac{1}{t})A(p_1, p_2, p_3, p_4),
}{}
and so on.  Inserting these geometric invariants into \Eq{amplitudes}, we obtain exactly the four- and five-point amplitudes input into the metric at the start. Note that the extension of this claim to six-point and beyond is more subtle, since it depends on the precise off-shell continuation of the four-point amplitude.





\Section{Gauge Theory and Gravity.} Our approach generalizes to any theory of massless bosons of arbitrary spin.  
For example, consider the geometry-kinematics replacement that maps the scalar of the NLSM to the gauge boson of YM theory,
$\phi^i   \rightarrow  A_\mu^a(p)$,
together with 
\eq{
g_{ij} \quad &\rightarrow \quad g^{\mu\nu}_{ab}(p_1, p_2) \delta(p_{12})\\
g_{ij,k} \quad &\rightarrow \quad g^{\mu\nu|\rho}_{ab|c}(p_1, p_2 |p_3) \delta(p_{123})\\
g_{ij,kl} \quad &\rightarrow \quad g^{\mu\nu|\rho\sigma}_{ab|cd}(p_1, p_2| p_3,p_4)  \delta(p_{1234}), 
}{map_metric_gauge}
and so on.  The only restriction on the form factors on the right-hand side is that they are symmetric under swapping the momenta simultaneously with the color and Lorentz indices for legs 1 and 2, and similarly for legs $3,4,\cdots, n$.  Working in Feynman gauge, the kinetic term fixes
$g_{ab}^{\mu\nu}(p_1, p_2) = \delta_{ab} \eta^{\mu\nu}$.
Meanwhile, the internal sums map to
$\sum_i    \rightarrow  \sum_a \sum_\mu \int_{p}$,
and so the field with lowered indices is simply
$\phi_i \rightarrow A^\mu_a(-p)$.
Applying this geometry-kinematics replacement to the NLSM Lagrangian in \Eq{L_NLSM},
we obtain the action  
\begin{widetext}
\eq{
S = \tfrac{1}{2} \int\displaylimits_{p_1, p_2}  (p_1 \cdot p_2) A_\mu^a(p_1) A_\nu^b(p_2) \bigg[  & \delta_{ab} \eta^{\mu\nu} \delta(p_{12}) + \int\displaylimits_{p_3}  g^{\mu\nu|\rho}_{ab|c}(p_1, p_2| p_3) \delta(p_{123})  A^c_\rho(p_3)  +\cdots \bigg].
}{action_gauge}
\end{widetext}
Meanwhile, YM theory in Feynman gauge is defined by
\eq{
L = -\tfrac{1}{4} F_{\mu\nu}^a F^{\mu\nu}_a - \tfrac{1}{2} \partial A^a \partial A_a,
}{}
which corresponds to \Eq{action_gauge} for the choice of metric
\begin{align}
\label{metric_YM}
 \quad g^{\mu\nu|\rho}_{ab|c}(p_1,p_2|p_3) &=   -\tfrac{i  f_{abc}}{p_1\cdot p_2}  (p_3^\mu \eta^{\nu\rho} -p_3^\nu \eta^{\mu\rho})  \\ \nonumber
  \quad g^{\mu\nu|\rho\sigma}_{ab|cd}(p_1,p_2|p_3, p_4) &=   - \tfrac{f_{abe} f_{cd}^{\phantom{cd}e}}{2 (p_1\cdot p_2)}  (\eta^{\mu\rho} \eta^{\nu\sigma}-\eta^{\mu\sigma} \eta^{\nu\rho}).
\end{align}
Computing kinematic curvature invariants and inserting them into the four- and five-point NLSM amplitudes in \Eq{amplitudes}, we correctly obtain those of YM theory.  

While the resulting amplitudes are of course gauge invariant, the kinematic curvature invariants which enter into them are not. This is unsurprising, since a generic diagrammatic expansion of the amplitude---for example the usual color-kinematics decomposition into kinematic numerators---is also not in general manifestly gauge invariant. This implies that to the extent to which the kinematic metric parameterizes a bona fide, underlying manifold, gauge field configurations that are gauge equivalent in fact label distinct points with different local curvatures.  
We leave for future study whether there exists a choice of kinematic metric that is gauge invariant.

The above procedure also applies to gravity.  We simply promote the scalar of the NLSM to the graviton via $\phi^i \rightarrow h_{\mu\nu}(p)$, with the kinematic metric given by
\eq{
g_{ij} \quad &\rightarrow \quad g^{\mu\nu\rho\sigma}(p_1, p_2) \delta(p_{12})\\
g_{ij,k} \quad &\rightarrow \quad g^{\mu\nu\rho\sigma|\alpha\beta}(p_1, p_2 |p_3) \delta(p_{123})\\
g_{ij,kl} \quad &\rightarrow \quad g^{\mu\nu\rho\sigma|\alpha\beta\gamma\delta}(p_1, p_2| p_3,p_4)  \delta(p_{1234}), 
}{map_metric_gravity}
and so on.  Here it is convenient to choose deDonder gauge, where the kinetic term, {\it i.e.}~inverse propagator numerator, is  $g^{\mu\nu\rho\sigma}(p_1,p_2)  = \tfrac{1}{2}(\eta^{\mu\rho}\eta^{\nu\sigma}+\eta^{\mu\sigma}\eta^{\nu\rho}-\eta^{\mu\nu}\eta^{\rho\sigma}) $.  Meanwhile, internal summations are mapped to $\sum_i  \rightarrow  \sum_\mu \sum_\nu \int_{p}$, while lowering indices corresponds to sending incoming momenta to outgoing while contracting with a tensor, $\phi_i \rightarrow g^{\mu\nu \rho\sigma} h_{\rho\sigma}(-p)$.  Applying this geometry-kinematics replacement to \Eq{L_NLSM}, one obtains the obvious generalization of \Eq{action} and \Eq{action_gauge} to gravity.

Our construction applies even more generally to any theory of multiple boson species of diverse flavors, spins, and interactions.  In particular, all tensorial manipulations in the NLSM transfer wholesale to a general theory, {\it i.e.}~the internal index $i$ is promoted to an array of quantum numbers labeling momenta, color, and spin Lorentz indices $(p,a, \mu, \cdots)$.  An explicit kinematic metric can always be computed directly from the Feynman vertices in the obvious generalization of \Eq{metric_from_Feynman}.  Thus, a geometric organization of the amplitudes can be achieved for a general theory of massless bosons, as we will elaborate on further in \cite{GKlong}.

\Section{Soft Theorems.} On-shell scattering amplitudes in the NLSM obey a universal soft theorem based on the geometry of field space \cite{Cheung:2021yog}.  The soft limit of the tree-level $n$-point amplitude is related to $(n-1)$-point via
\eq{
&\lim\limits_{p_{n}\rightarrow0}  A_{i_1\cdots i_{n}}  
= \nabla_{i_n} A_{i_1 \cdots i_{n-1}} \\
&\qquad = \partial_{i_n} A_{i_1 \cdots i_{n-1}}- \sum_{\alpha=1}^{n-1}\Gamma^{j}_{\phantom{j} i_n j_\alpha}A_{i_1 \cdots j\cdots i_{n-1}} ,
}{geosofteq}
hereafter dropping all terms of order ${\cal O}(p_n)$.  Note that the partial derivative acts with respect to the vacuum expectation value of the field, so $\partial_{i_n} = \tfrac{\partial}{\partial \langle \phi^{i_n}\rangle}$.  \Eq{geosofteq} has the  physical interpretation that the soft limit simply shifts the vacuum expectation value of the corresponding background scalar field. Note that \Eq{geosofteq} automatically incorporates any cubic potential potential terms implicitly through the covariant derivative, provided the potential is included in the metric as in \Eq{metric_V}.

Now consider the Lagrangian for scalar quantum electrodynamics, $L = -|D\phi|^2$ where $D_\mu  = \partial_\mu - i A_\mu $.  The photon couples to a pair of complex scalars via the metric
\eq{
g^\rho_{\bar \phi  \phi}(p_1,p_2|p_3)  = - g^\rho_{ \phi \bar \phi }(p_1,p_2|p_3)  = \tfrac{1}{p_1\cdot p_2} (p_1-p_2)^\rho .
}{}
We now compute the corresponding kinematic connection and insert it into \Eq{geosofteq}, taking the leg $n$ to be a photon and legs $1,\cdots, n-1$ to be scalars.   For example, this maps
\eq{
-\Gamma^{j}_{\phantom{j} i_n j_\alpha}A_{i_1 \cdots j\cdots i_{n-1}} \quad \rightarrow  \quad & 
q_\alpha \tfrac{ e_n\cdot  p_\alpha}{p_n \cdot p_\alpha}  (1+ p_n \cdot \tfrac{\partial}{\partial p_\alpha} ) A,
 }{}
 where $e_n$ is the polarization vector of the soft particle and $p_\alpha$ and $q_\alpha=\pm 1$ are the momentum and charge of the hard leg. Here $p_n \cdot \tfrac{\partial}{\partial p_\alpha} $ appears because the contraction with the connection shifts the momentum $p_\alpha$ in the lower-point amplitude  to $p_\alpha + p_n$.
Meanwhile, the partial derivative with respect to a background scalar in \Eq{geosofteq} maps to one with respect to a background photon,
\eq{
\tfrac{\partial}{\partial \langle \phi^{i_n}\rangle} \quad \rightarrow \quad e^{\mu}_n \tfrac{\partial}{\partial \langle A_{\mu}\rangle} =  -\sum_{\alpha=1}^{n-1} q_\alpha e_n  \cdot   \tfrac{\partial}{\partial p_\alpha},
}{}
where we have used that the background photon is a connection, and thus couples universally to momenta.  Thus,  the covariant derivative of the NLSM maps to the leading and subleading soft photon theorems \cite{Low:1958sn, Weinberg:1964ew, Weinberg:1965nx,  Burnett:1967km},
\eq{
\nabla_i \quad \rightarrow\quad \sum_{\alpha=1}^{n-1} \tfrac{ 1}{p_n \cdot p_\alpha}   \left[ e_n\cdot  p_\alpha +  e_n \cdot J_\alpha \cdot p_n \right],
}{}
where $J_\alpha^{\mu\nu} = p_\alpha^\mu \tfrac{\partial}{\partial p_{\alpha \nu}}-p_\alpha^\nu \tfrac{\partial}{\partial p_{\alpha \mu}}$ is the angular momentum operator acting on a hard leg. The generalization to soft gluons \cite{Berends:1988zn, Casali:2014xpa, Bern:2014vva} and gravitons \cite{Weinberg:1964ew, Weinberg:1965nx, Gross:1968in, White:2011yy, Cachazo:2014fwa, Bern:2014vva} will be explored in detail in \cite{GKlong}.

\Section{Conclusions.} We have shown that an arbitrary theory of massless bosons is classically equivalent to an NLSM with an internal field space metric which is momentum-dependent.  The resulting kinematic metric induces a corresponding geometrical structure, wherein the associated kinematic curvatures are tensors under general field redefinitions.  Armed with these geometric objects, we trivially obtain any tree amplitude of massless bosons from those of the NLSM by a replacement of geometry for kinematics.

Our analysis has focused solely on tree-level dynamics, but geometry-kinematics duality should also apply to loops in the very same way as does color-kinematics duality.  In particular, any loop integrand constructed via generalized unitarity from tree amplitudes will be manifestly a function of loop momentum-dependent kinematic curvature invariants.  Upon integration, the resulting amplitudes will agree with standard methods.  However, like in color-kinematics duality, the geometry-kinematics replacement must be applied at integrand level and not post integration.

\medskip

\noindent {\it Note added:} During the completion of this paper, we learned of interesting concurrent work by Craig, Cohen, Lu, and Sutherland \cite{CCLS}, which proposes a related but different notion of functional geometry for effective field theories. We are grateful to those authors for sharing their draft with us.

\Section{Acknowledgments.}
C.C., A.H., and J.P.-M. are supported by the DOE under grant no. DE- SC0011632 and by the Walter Burke Institute for Theoretical Physics.  We are grateful to Zvi Bern, Enrico Herrmann, James Mangan, Aneesh Manohar, and Ira Rothstein for comments on the draft.

\bibliographystyle{utphys-modified}
\bibliography{bibliographyGK}

\end{document}